\documentclass[aip,reprint]{revtex4-1}
\usepackage{graphicx,amsmath,amssymb,color,amsthm}
\usepackage{bigints}
\usepackage{relsize}
\graphicspath{ {figures/} }

\begin{document}

\title{Performance of chaos diagnostics based on Lagrangian descriptors. Application to the 4D standard map }

\author{S.~Zimper}
\affiliation{Nonlinear Dynamics and Chaos Group, Department of Mathematics and
Applied Mathematics, University of Cape Town, Rondebosch 7701,
South Africa.}

\author{A.~Ngapasare}
\affiliation{Nonlinear Dynamics and Chaos Group, Department of Mathematics and
Applied Mathematics, University of Cape Town, Rondebosch 7701,
South Africa.}

\author{M.~Hillebrand}
\affiliation{Nonlinear Dynamics and Chaos Group, Department of Mathematics and
Applied Mathematics, University of Cape Town, Rondebosch 7701,
South Africa.}
\affiliation{Max Planck Institute for the Physics of Complex Systems, Nöthnitzer Stra\ss e 38, Dresden, 01187, Germany}
\affiliation{ Center for Systems Biology Dresden, Pfotenhauer Stra\ss e 108, Dresden, 01307, Germany}
    
\author{M.~Katsanikas}
\affiliation{Research Center for Astronomy and Applied Mathematics, Academy of Athens, Soranou Efesiou 4, Athens, GR-11527, Greece.}

\author{S.~Wiggins}
\affiliation{School of Mathematics, University of Bristol, Fry Building, Woodland Road, Bristol, BS8 1UG, United Kingdom.}
\affiliation{Department of Mathematics, 
United States Naval Academy, 
Chauvenet Hall, 572C Holloway Road, 
Annapolis, MD 21402-5002. 
}

\author{Ch.~Skokos}
\email[]{haris.skokos@uct.ac.za}
\affiliation{Nonlinear Dynamics and Chaos Group, Department of Mathematics and
Applied Mathematics, University of Cape Town, Rondebosch 7701,
South Africa.}


\date{\today}

\begin{abstract}
We investigate the ability of simple diagnostics based on Lagrangian descriptor (LD) computations of initially nearby orbits to detect chaos in conservative dynamical systems with phase space dimensionality higher than two. In particular, we consider the recently introduced methods of the difference ($D_L^n$) and the ratio ($R_L^n$) of the LDs of neighboring orbits, as well as a quantity ($S_L^n$) related to the finite-difference second spatial derivative of the LDs, and use them to determine the chaotic or regular nature of ensembles of orbits of a prototypical area-preserving map model, the 4-dimensional (4D) symplectic standard map. Using the distributions of the indices’ values we determine appropriate thresholds to discriminate between regular and chaotic orbits, and compare the obtained characterization against that achieved by the Smaller Alignment Index (SALI) method of chaos detection, by recording the percentage agreement  $P_A$ between the two classifications. We study the influence of various factors on the performance of these indices, and show that the increase of the final number of orbit iterations $T$ and the order $n$ of the indices (i.e.~the dimensionality of the space where the considered nearby orbits lie), as well as the decrease of the distance $\sigma$ of neighboring orbits, increase the $P_A$ values along with the required computational effort. Balancing between these two factors we find appropriate $T$, $n$ and $\sigma$ values, which allow the efficient use of the $D_L^n$, $R_L^n$ and $S_L^n$ indices as short time and computationally cheap chaos diagnostics achieving $P_A \gtrsim 90\%$, with $D_L^n$ and $S_L^n$ having larger $P_A$ values than $R_L^n$. Our results show that the three LDs-based indices perform better for systems with large percentages of chaotic orbits. In addition, our findings clearly indicate the capability of LDs to  efficiently identify chaos in systems whose phase space is difficult to visualize (due to its high dimensionality), without knowing  the variational equations (tangent map) of continuous (discrete) time systems needed by traditional chaos indicators. 
\end{abstract}


\maketitle







\section{Introduction}
\label{sec:introduction}

Determining the nature of individual orbits as either chaotic or regular, as well as the dynamics of ensembles of orbits, is fundamental for understanding the behavior of continuous and discrete time dynamical systems. To this end, a variety of different techniques and indicators, to either visualize the system's phase space or to detect chaotic orbits, have been developed over the course of time. 

The asymptotic measures introduced by Lyapunov~\cite{lyapunov1992} to characterize the growth or shrinking of small phase space perturbations to orbits (often referred to as deviation vectors)  have been widely accepted as a standard tool for this purpose. These quantities are commonly named Lyapunov exponents (LEs). Following the formulation of the  multiplicative ergodic theorem by  Oseledec~\cite{oseledec1968}, a theoretical basis for the numerical computation of LEs was presented~\cite{benettin1980,benettin1980b}. The estimation of the maximum LE (mLE) through the numerical computation of the finite-time mLE (ftmLE), is nowadays one of the most commonly used chaos detection methods as the positivity of the mLE of bounded orbits, which do not escape to infinity, indicates chaotic behavior (see for example~\cite{S10} and references therein). 

The slow convergence of the ftmLE to its limiting value has necessitated the search for alternative, more efficient indicators. Among these indicators are the so-called fast Lyapunov Indicator (FLI)~\cite{froeschle1997} and its variants~\cite{B05}, the Mean Exponential Growth of Nearby Orbits (MEGNO)~\cite{CS00}, the Smaller Alignment Index (SALI)~\cite{skokos2001} and its extension, the Generalized Alignment Index (GALI)~\cite{skokos2007}. These indicators have certain advantages over the estimation of the mLE as, in general, they manage to characterize orbits as regular or chaotic faster and with less computational effort,  although they also rely on the time evolution of at least one deviation vector.

One of the most successful methods among this set of new indicators is the SALI, which has been efficiently used to  study of chaoticity of several different systems, such as accelerator models~\cite{Petalas2008,BCSV12}, predator-prey population maps~\cite{Saha2012}, Bose-Einstein condensates~\cite{KKSK14}, galactic potentials~\cite{MBS13,Carpintero2014}, as well as nuclear physics models~\cite{Stransky2009}. The interested reader is referred to the review~\cite{SM16} for more details on this method and its applications.

A recently developed visualization technique for the identification of phase space structures in continuous time dynamical systems and discrete time iterative maps is the method of Lagrangian descriptors (LDs)~\cite{madrid2009,mancho2013,ldbook2020}. The computation of LDs is based on the accumulation of some positive scalar value along the path of individual orbits to produce a scalar field on a  grid of initial conditions. From the gradient of this field the manifolds in both regular and chaotic regions can be identified as singular features, following the theoretical discussions in~\cite{lopesino2015} and~\cite{lopesino2017} for the discrete and continuous time settings respectively. Initially applied to the study of ocean currents~\cite{madrid2009,mancho2013}, this method has since been utilized to study the dynamics of systems from a variety of different fields such as chemical transition state theory \cite{Craven2015,craven2017}, molecular systems \cite{revuelta2019,KHSW23}, cardiovascular flows \cite{darwish2021}, and stochastic dynamical systems~\cite{balibrea2016}. 

In~\cite{montes2021} the characterization of regular motion by LDs was considered, while in a recent work~\cite{Daquin2022} an indicator based on the estimation of the second derivative of the LDs field was used to discriminate between regular and chaotic motion in discrete and continuous systems. These works paved the way for LDs to be used for not only a visual inspection of the phase space, but also for   determining the chaotic nature of orbits. This was  done in~\cite{Hillebrandchaos2022}, where it was shown that indicators derived from LDs of nearby orbits can be used to characterize the chaoticity of ensembles of orbits with $\gtrsim90\%$ accuracy (in comparison with the characterizations obtained by the SALI method) for both the H\'enon--Heiles~\cite{henon1964} system and the two-dimensional ($2$D) standard map~\cite{chirikov1979}. An advantage of LDs-based chaos diagnostics over the more traditional above mentioned chaos indicators is that the evolution of deviation vectors is not required, which reduces  the  complexity of the performed computations  and simultaneously diminishes the required CPU time. 

In~\cite{Hillebrandchaos2022} the introduced methods were applied to low-dimensional systems having 2D phase spaces, which are easily depicted. Here we extend that study by investigating in detail the performance of these diagnostics in a higher-dimensional setting, where the phase space's visualization becomes challenging, although methods like the `color and rotation'~\cite{katsanikas2013,zachilas2013structure} and the `phase space slices'~\cite{richter2014}, as well as approaches based on LDs~\cite{agaoglou2021visualizing} have been used for that purpose. In particular, we demonstrate how these techniques can be used to identify orbits as regular or chaotic within a certain accuracy, using as a test case a $4$D  map, a higher-dimensional conservative dynamical system of discrete time.

The rest of the paper is organized as follows. In Sect.~\ref{sec:theory} we describe the numerical computation of the various chaos diagnostics used in this investigation. In Sect.~\ref{sec:numerica_results}, we implement our techniques for studying the chaotic behavior of the 4D standard map for different setups of the system. Finally in Sect.~\ref{sec:conclusions}, we discuss our findings and summarize our conclusions.

\section{Numerical techniques} 
\label{sec:theory}

In order to study the performance and efficiency of the three quantities based on the LDs values of neighboring orbits, which were  presented in~\cite{Hillebrandchaos2022},  for systems of higher dimensionality we consider here, as a test case of an area preserving map, the $4$D standard map~\cite{KG8} obtained by coupling two (identical in our implementation) $2$D standard maps
\begin{equation}
    \label{eq:4Dmap}
    \begin{aligned}
    x_1' & = x_1 + x_2 ',   \\
    x_2' & = x_2 + \frac{K}{2 \pi}\sin(2 \pi x_1)  - \frac{B}{2 \pi}\sin \big[ 2 \pi(x_3 - x_1) \big] , \\
    x_3' & = x_3 + x_4 '  , \\
    x_4' & = x_4 + \frac{K}{2 \pi}\sin(2 \pi x_3)  - \frac{B}{2 \pi}\sin \big[2 \pi(x_1 - x_3) \big], 
    \end{aligned} (\mbox{mod} \, 1)
\end{equation}
with $K$ and $B$ being real parameters, and $ \mathbf{z}=( x_1' , x_2' , x_3' , x_4' )$ denoting the state vector of the map's coordinates after a single iteration.  The parameter $K$ defines  the nonlinearity strength  of each one of the $2$D coupled maps, while $B$ determines  the strength of  coupling  between the two $2$D maps. All coordinates are given (mod 1), so that $0 \leq x_i <1$, $i=1,2,3,4$. We note that the number $T$  of map's iterations will also be referred to as the (discrete) time of the system.

Small perturbations of tested orbits are key in determining the regular or chaotic nature of these orbits. Such a perturbation defines the  deviation vector $ \mathbf{w}=(\delta x_1, \delta x_2, \delta x_3, \delta x_4 ) $, whose time evolution is governed by the system’s tangent map given by 
\begin{equation}
   \label{eq:4Dmap_var}
  \begin{aligned}
    \delta x_1 ' = \, &  \delta x_1 + \delta x_2 ', \\
    \delta x_2 ' = \, &  \Big\{ K \cos(2 \pi x_1) + B \cos\big[2 \pi (x_3 - x_1)\big] \Big\} \delta x_1  \\
      & + \delta x_2 - B \cos\big[ 2 \pi (x_3 - x_1) \big] \delta x_3, \\
    \delta x_3 ' = \, &  \delta x_3 + \delta x_4 ', \\
    \delta x_4 ' = \, &  -B \cos \big[2 \pi (x_1 - x_3)\big] \delta x_1 \\
     & +\Big\{ K \cos(2 \pi x_3) + B \cos \big[ 2 \pi (x_1 - x_3) \big] \Big\} \delta x_3 + \delta x_4.     
  \end{aligned}
\end{equation}

The mLE $ \lambda_1 $ of an orbit is estimated through the computation of the ftmLE
\begin{equation} 
\label{eq:mLE} 
\Lambda (T) = \frac{1}{T} \ln \left( \frac{\| \mathbf{w}(T)\| }{\| \mathbf{w}(0) \| }  \right),
\end{equation}
as 
\begin{equation}
\lambda_1 = \lim_{T \to \infty } \Lambda (T), 
\end{equation}
with $\| \cdot \|$ denoting the usual Euclidean norm of a vector. For a chaotic orbit, $ \Lambda$  eventually saturates to a positive value, whereas in the case of regular orbits $ \Lambda$ decreases following the power law~\cite{S10} 
\begin{equation} 
\label{eq:mLE_reg} 
\Lambda (T) \propto \ln (T)/ T.
\end{equation}

In contrast to the estimation of the mLE, the computation of the SALI depends on the evolution of two, initially linearly independent, 
deviation vectors $\mathbf{w}_1$ and $\mathbf{w}_2$. Then SALI$(T)$, which quantifies the alignment of these two deviation vectors, is computed as
\begin{equation} 
\label{eq:SALI} 
\mbox{SALI}(T)=\min \big\{\|
    \hat{\mathbf{w}}_1(T) + \hat{\mathbf{w}}_2(T) \|, \| \hat{\mathbf{w}}_1(T)
    - \hat{\mathbf{w}}_2(T) \| \big\},
\end{equation}
with  
\begin{equation} 
\hat{\mathbf{w}}_k(T) = \frac{\mathbf{w}_k(T)}{\| \mathbf{w}_k(T) \|}, \qquad k=1,2,
\end{equation}
being a  vector of unit norm. For chaotic orbits, the two deviation vectors will eventually be aligned to the direction related to the mLE and consequently the SALI will follow an exponential decay to zero, with a rate depending on the values of the two largest LEs $\lambda_1 \geq \lambda_2$. On the other hand, for regular orbits in the phase space of a $4$D symplectic map the SALI remains positive and practically constant. Thus, in summary, the behavior of the SALI for orbits of the $4$D standard map \eqref{eq:4Dmap} is 
\begin{equation}
\label{eq:SALI_4Dmap}
\mbox{SALI}(T) \propto
\begin{cases}
\mbox{constant} & \mbox{for regular orbits} \\
e^{-(\lambda_1-\lambda_2) T} & \mbox{for chaotic orbits}
\end{cases}. 
\end{equation}

In our study, following~\cite{Hillebrandchaos2022}, we  exploit the ability of LDs to capture the basic dynamical features  of a system in order to identify regular and chaotic motion. Let us first recall that the  ``$p$-norm'' definition of the LD for a discrete map is given by
\begin{equation}
    \label{eq:defDisLD}
    LD = \sum_{j=-T}^{T-1} \sum_{i=1}^{N}\left|z^{(i)}_{j+1}-z^{(i)}_{j}\right|^p, \qquad 0< p \leq 1,
\end{equation}
where $i$ indexes the $N$ elements of the state vector $\mathbf{z}$ [for map~\eqref{eq:4Dmap} $N=4$], and $j$ counts the map's iterations. Since the $LD$ definition \eqref{eq:defDisLD} with $p=0.5$ has been  successfully implemented in various studies (e.g.,~\cite{demian2017,Katsanikas2020b})  and  has shown a remarkable ability in identifying phase space structures, we will also set $p=0.5$ for our investigations. We emphasize that, although in the formal definition~\eqref{eq:defDisLD} of the LD the map is integrated both in the past ($j$ starts at $j=-T$) and in the future ($j$ goes up to $j=T-1$), for the purposes of our study the computation of the LDs through only forward (or backward) iterations is sufficient. More specifically, the presented results are solely obtained through forward iterations.   

Let us now discuss how we can identify the regular or chaotic nature of an orbit with initial conditions (ICs) at point $\mathbf{z}$ in the map's phase space, based on the values of the LDs of this orbit and of initially neighboring ones. The ICs of these neighboring orbits can be seen as grid points of a mesh in several spatial dimensions. In the case of the $4$D map \eqref{eq:4Dmap} we can consider neighboring orbits to an IC in $n$D spaces with $1 \leq n \leq 4$. For $n=1$ we have two neighboring points of $\mathbf{z}$ on a line (1D space), while for $n=2$ the four nearest neighbors are located on a grid in a 2D subspace of the 4D phase space. Thus, considering ICs of orbits on a finite grid of an $n(\geq 1)$D subspace of the $N(\geq n)$D phase space of a general $N$D symplectic map, any non-boundary grid point $\mathbf{z}$ in this subspace has $2n$ nearest neighbors
\begin{equation}
\label{eq:neighbors}
     \mathbf{y}_i^{\pm} = \mathbf{z}\pm \sigma^{(i)} \mathbf{e}^{(i)}, \,\, i=1,2,\ldots n,
 \end{equation} 
where $\mathbf{e}^{(i)}$  is the $i$th unit vector of the usual basis in $\mathbb{R}^n$, and $\sigma^{(i)}$ is the distance between successive grid points in this direction. 

If we respectively denote by $ LD(\mathbf{z})$ and $ LD\left(\mathbf{y}_i^\pm\right)$ the LDs of orbits with ICs $\mathbf{z}$ and $\mathbf{y}_i^\pm$ we can define the three diagnostics we use in our study,  following~\cite{Hillebrandchaos2022}. More specifically, the difference $D_L^n$ of LDs of neighboring orbits at $\mathbf{z}$ in an $n$D subspace is defined as 
\begin{equation}
     \label{eq:defDNLD}
     \displaystyle D_L^n(\mathbf{z}) = \frac{1}{2n} \sum_{i=1}^{n} \frac{\left| LD(\mathbf{z})- LD(\mathbf{y}_i^+)\right| +\left| LD(\mathbf{z}) - LD(\mathbf{y}_i^-)\right|}{ LD(\mathbf{z})},
\end{equation}
while the  ratio $R_L^n$ is given by  
\begin{equation}
     \label{eq:defRNLD}
     R_L^n(\mathbf{z}) = \left|1-\frac{1}{2n}\sum_{i=1}^{n}\frac{LD(\mathbf{y}_i^+) +LD(\mathbf{y}_i^-)}{{LD(\mathbf{z})}}\right|,
\end{equation}
with $n$ also referred to as the order of the index. The last indicator we use is related to the second spatial derivative of the LD quantity. It was introduced in~\cite{Daquin2022},  briefly studied in~\cite{Hillebrandchaos2022}, and applied to celestial mechanics problems in~\cite{DC22}, where it was denoted by the rather cumbersome notation $||\Delta LD||$. Here we adopt the notation $S_L^n$ to follow similar conventions to the notations of Eqs.~(\ref{eq:defDNLD}) and (\ref{eq:defRNLD}), as well as to clearly indicate the dimensionality of the grid on which this diagnostic is computed, and define the order $n$ index as 
 \begin{equation}
 \label{eq:defDfNLD}
     S_L^n (\mathbf{z}) = \frac{1}{n}  \sum_{i=1}^{n} \left|\frac{ LD(\mathbf{y}_i^+)  - 2LD(\mathbf{z}) + LD(\mathbf{y}_i^-)}{(\sigma^{(i)})^2} \right| .
 \end{equation}
We note that a difference between the definition of $S_L^n$ and of the $||\Delta LD||$ index used in~\cite{Daquin2022,Hillebrandchaos2022,DC22} is that in \eqref{eq:defDfNLD} the factor $1/n$ is introduced in order to compute a quantity `per dimension' of the space where the used ICs are, similar  to what is done in \eqref{eq:defDNLD} and \eqref{eq:defRNLD}.

\begin{figure*}
    \begin{center}
    \includegraphics[width=0.99\textwidth]{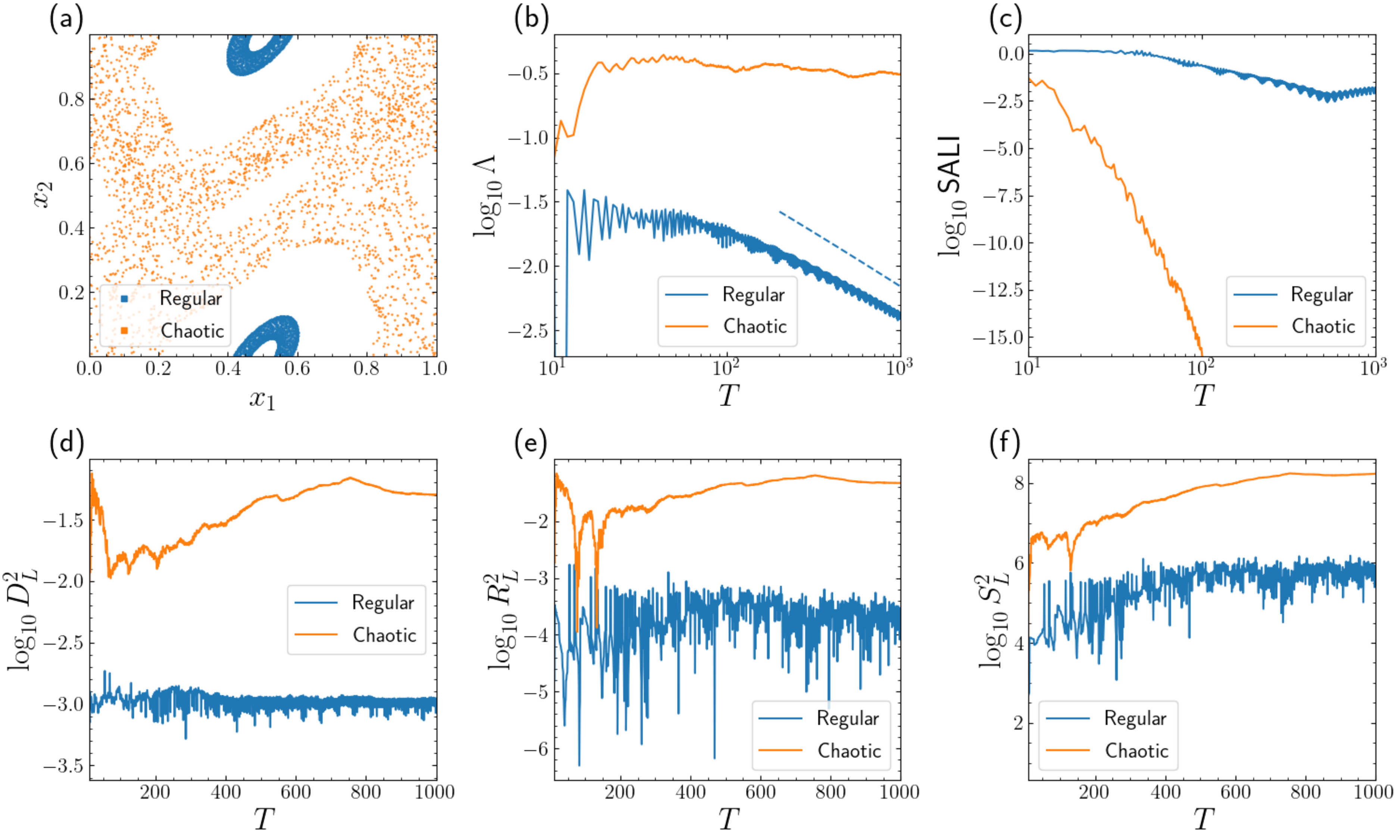}
    \caption{(a) The projection of a regular orbit (blue points)  with ICs $x_1=0.6$, $x_2=0.05$, $x_3=0.54$, $x_4=0.01$, and a chaotic orbit (orange points) with ICs $x_1=0.2$, $x_2=0.2$, $x_3=0.54$, $x_4=0.01$, of the $4$D  map \eqref{eq:4Dmap}  with $K=1.5$ and $B=0.05$ on the $(x_1, x_2)$ plane for $T=2500$ forwards iterations of the map. Time evolution of (b) the $\Lambda$ \eqref{eq:mLE}, (c) the SALI \eqref{eq:SALI}, (d) the $D_L^2$ \eqref{eq:defDNLD}, (e) the $R_L^2$ \eqref{eq:defRNLD}, and (f) the $S_L^2$ \eqref{eq:defDfNLD} for the two orbits of (a). The $D_L^2$, $R_L^2$ and $S_L^2$ are  evaluated in the plane  $(x_1, x_2)$,  with a grid spacing $\sigma=10^{-3}$ in each direction. The dashed line in (b) denotes the function $\ln(T)/T$ \eqref{eq:mLE_reg}.}
    \label{fig:reg_vs_chaotic_plots}
    \end{center}   
\end{figure*}

In order to demonstrate the basic behaviors of the two chaos indicators [$\Lambda$ \eqref{eq:mLE}, SALI \eqref{eq:SALI}] and the three LDs-based diagnostics [$D_L^n$ \eqref{eq:defDNLD}, $R_L^n$ \eqref{eq:defRNLD}, $S_L^n$ \eqref{eq:defDfNLD}] we use in our study,  we compute them for two representative orbits, one regular with ICs $x_1=0.6$, $x_2=0.05$, $x_3=0.54$, $x_4=0.01$ and one chaotic with ICs $x_1=0.2$, $x_2=0.2$, $x_3=0.54$, $x_4=0.01$, for the 4D map \eqref{eq:4Dmap} with $K=1.5$ and $B=0.05$. We note that for the three diagnostics based on LDs computations we set $n=2$, consider neighboring orbits on a square grid in the $(x_1, x_2)$ plane and compute the order 2 version of the indices. The projection of the $T=2500$ consequents of the regular (blue points) and the chaotic orbit (orange points) on the plane  $(x_1, x_2)$  are shown in Fig.~\ref{fig:reg_vs_chaotic_plots}(a). The points of the regular orbit lie on a 4D stability island and create a regular, torus-like structure. On the other hand, the consequents of the chaotic orbit correspond to the scattered point in Fig.~\ref{fig:reg_vs_chaotic_plots}(a).

In Figs.~\ref{fig:reg_vs_chaotic_plots}(b)-(f) we respectively plot the time evolution of $\Lambda$, SALI, $D_L^2$, $R_L^2$ and $S_L^2$ for the considered regular (blue curves) and chaotic orbit (orange curves). From the results of Fig.~\ref{fig:reg_vs_chaotic_plots}(b) we see that the ftmLE $\Lambda$ of the regular orbit eventually decreases to zero proportionally to $\ln(T)/T$ (dashed line), while for  the chaotic orbit it saturates to a positive value as expected. On the other hand, the  SALI [Fig.~\ref{fig:reg_vs_chaotic_plots}(c)] approaches a positive value for the regular orbit while it tends exponentially fast to zero  for the chaotic one. We note that all computations throughout this study are performed using double-precision accuracy, thus we stop the time evolution of the  SALI when its values reach $10^{-16}$, i.e.~the  machine precision. From Fig.~\ref{fig:reg_vs_chaotic_plots}(c) we see that the SALI of the chaotic orbit requires only about $T=100$ forwards iterations to reach the $10^{-16}$ threshold, characterizing the orbit beyond any doubt as chaotic as its SALI is practically zero.

From the results presented in Figs.~\ref{fig:reg_vs_chaotic_plots}(d)-(f) we see that the values of $D_L^2$, $R_L^2$ and $S_L^2$ of the regular orbit remain well above the ones obtained for the chaotic one (apart from some short initial time interval $T\lesssim 200$ for $R_L^2$). These clear differences between the values of the $D_L^2$, $R_L^2$ and $S_L^2$ diagnostics for regular and chaotic orbits are observed generally and are not related to the particular example orbits shown here. Thus, as was presented in~\cite{Hillebrandchaos2022}, and will be discussed in detail in Sect.~\ref{sec:numerica_results}, we can define appropriate threshold values for each one of these three diagnostics to efficiently discriminate between regular and chaotic orbits. Nevertheless, it is important to note that this distinction needs a minimum (rather small) number of iterations in order to be clearly established, as we see in Figs.~\ref{fig:reg_vs_chaotic_plots}(e) and (f).

\section{Numerical Results}
\label{sec:numerica_results}

In this section we investigate in detail the ability of the $D_L^n$, $R_L^n$ and $S_L^n$ indices to distinguish between regular and chaotic orbits in dynamical systems whose phase space dimension is higher than two. As a representative case of such a system we consider the prototypical $4$D standard map \eqref{eq:4Dmap}. In our study we investigate the influence of various factors on the ability of the indicators to accurately characterize the chaoticity of orbits, like the number of the performed map iterations, the extent of the system's chaoticity (i.e.~the fraction of the chaotic orbits), and  the order of the indicators.

\subsection{Dynamics on a $2$D subspace}
\label{sec:2D_plane}

\begin{figure*}
    \begin{center}
    \includegraphics[width=0.99\textwidth]{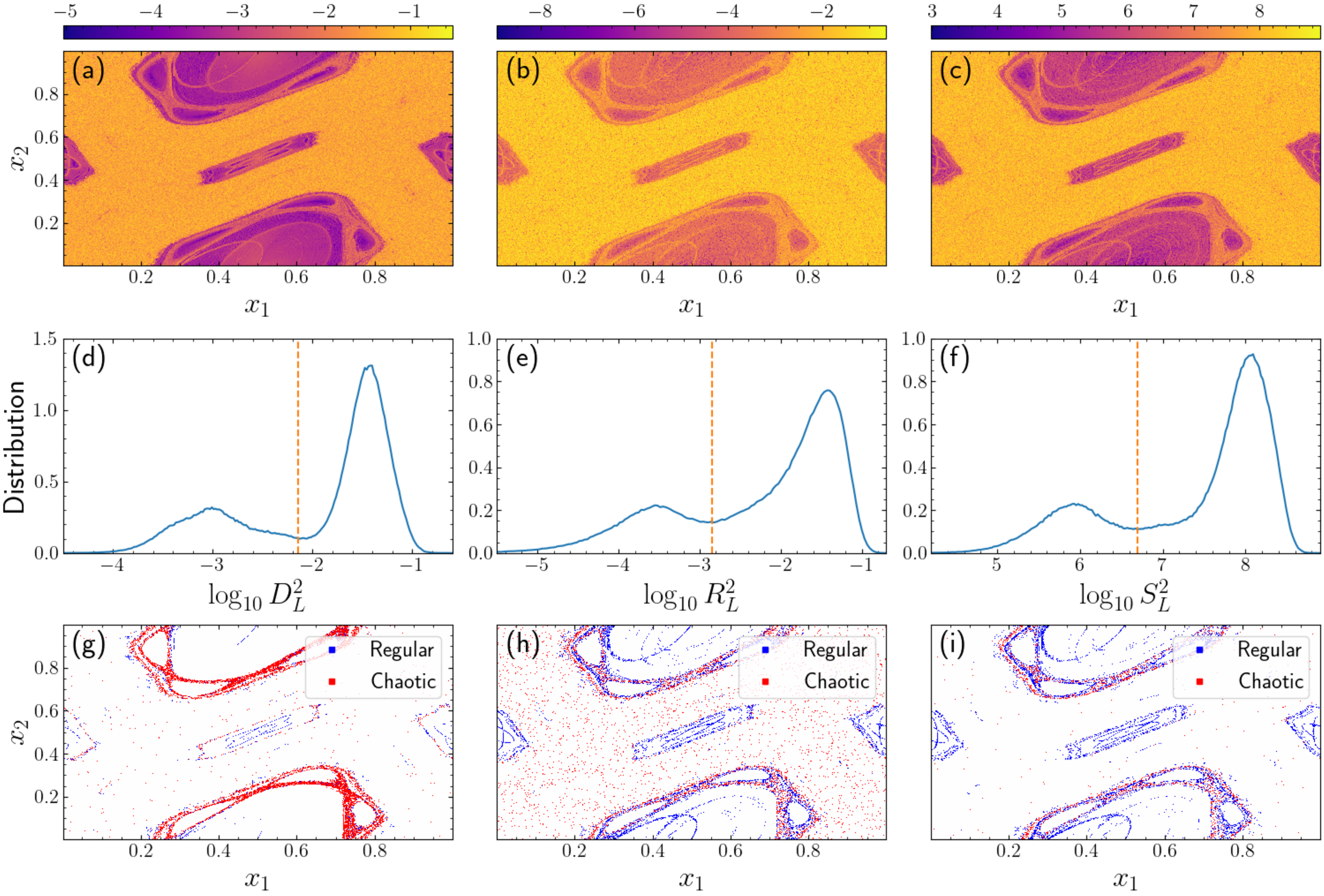}
    \caption{Results obtained for orbits having their ICs on a $1000 \times 1000$ grid on the $2$D subspace $(x_1, x_2)$ with $x_3 = 0.54$, $x_4 = 0.01$, of the $4$D map \eqref{eq:4Dmap}  for $K=1.5$ and $B=0.05$, after $T=10^3$  forward iterations. The ICs are colored according to the orbits' (a) $\log_{10} D_L^2$ (\ref{eq:defDNLD}),   (b) $\log_{10} R_L^2$ (\ref{eq:defRNLD}), and (c) $\log_{10} S_L^2$ (\ref{eq:defDfNLD}) values, using the color scales at the top of each panel. Normalized distributions of the (d)  $\log_{10} D_L^2$, (e)  $\log_{10} R_L^2$  and (f)  $\log_{10} S_L^2$ values of the orbits considered in (a)-(c). The values $\log_{10} D_L^2 = -2.14$, $\log_{10} R_L^2 = -2.85$ and $\log_{10} S_L^2 = 6.70$ are respectively denoted in (d), (e) and (f) by an orange vertical, dashed line. The set of the considered ICs  which are incorrectly characterized by, (g) the $D_L^2$, (h) the $R_L^2$, and (i) the $S_L^2$ index, with blue points corresponding to regular orbits (according to the classification obtained by the SALI method for $T=10^3$) which are falsely identified as chaotic, and red points denoting  chaotic orbits which are incorrectly identified as regular.}
    \label{fig:NLD_K1.5_colour}        
    \end{center}
\end{figure*}

Extending the results presented in~\cite{Hillebrandchaos2022} for dynamical systems with $2$D phase spaces (in particular the H{\'e}non-Heiles Hamiltonian~\cite{henon1964} and the $2$D standard map \cite{chirikov1979}) to the $4$D map \eqref{eq:4Dmap}, we  first investigate the performance of the $D_L^n$, $R_L^n$ and $S_L^n$ indices in a $2$D subspace of the map for which we can easily obtain the direct visualization of regular and chaotic regions. In particular, we consider a grid of $1000 \times 1000$ equally spaced ICs in the subspace $(x_1, x_2)$ by setting $x_3 = 0.54$ and $x_4 = 0.01$, for  $K=1.5$ and $B= 0.05$. This arrangement sets  the distance between immediate neighboring ICs to $\sigma = 10^{-3}$ in both directions on the $(x_1, x_2)$ plane. The LDs of all these orbits are computed for $T=10^3$ forward iterations, and from the obtained results we evaluate indicators $D_L^n$, $R_L^n$  and $S_L^n$  for each IC. Since the considered ICs lie on a $2$D plane, we compute the order $n=2$ versions of the three indicators. In Figs.~\ref{fig:NLD_K1.5_colour}(a)-(c), we present the resulting color plots of these computations, where ICs on the $(x_1, x_2)$ plane are colored according to their $\log_{10} D_L^2$, $\log_{10} R_L^2$ and $\log_{10} S_L^2$ values respectively. These plots display similar characteristics, providing a clear qualitative description of the structure of the phase space, with regular regions (islands of stability) corresponding to areas of lower values and chaotic regions (chaotic sea) having higher values, in accordance to what was found in~\cite{Hillebrandchaos2022}.

Although the color plots in  Figs.~\ref{fig:NLD_K1.5_colour}(a)-(c) correctly capture the overall dynamical features of the system, our main goal is to use the three indices for obtaining a quantitative identification of orbits as regular or chaotic. In order to obtain an estimation of the chaos extent in the studied $2$D subspace of the map a threshold value needs to be established for each index, so that orbits can be characterized as chaotic or regular if they respectively result in index values above or below these thresholds.  In Figs.~\ref{fig:NLD_K1.5_colour}(d)-(f) we show the normalized distributions of the logarithms of these three quantities, all of which clearly show two peaks separated by a trough, which demarcates ICs leading to regular (low values) and chaotic motion (high values). Assuming that the minimum between the two peaks provides a good threshold value for discriminating between regular and chaotic orbits, the following values are obtained: $\log_{10} D_L^2 = -2.14$, $\log_{10} R_L^2 = -2.85$ and $\log_{10} S_L^2 = 6.70$, respectively denoted by orange vertical,  dashed lines in Figs.~\ref{fig:NLD_K1.5_colour}(d)-(f). 

As was also observed in~\cite{Hillebrandchaos2022} this approach does not necessarily lead to the correct characterization of all orbits, with discrepancies mainly appearing at the edges of stability islands.   We investigate if this trend also persists for the $4$D standard map by comparing the characterization obtained from the $D_L^2$, $R_L^2$ and $S_L^2$ diagnostics against the one made by the SALI indicator for $T=10^3$ iterations. Noting that the SALI of regular orbits will fluctuate around a positive, constant value, while for chaotic orbits it will exponentially decrease to zero [see Eq.~\eqref{eq:SALI_4Dmap} and Fig.~\ref{fig:reg_vs_chaotic_plots}(c)], we consider a threshold value of $\log_{10} \mbox{SALI} = -8$, so that an orbit is characterized as regular if $\log_{10} \mbox{SALI} \geq -8$, and as chaotic if $\log_{10} \mbox{SALI} < -8$. The percentage agreement $P_A$ of the characterization of the orbits of Fig.~\ref{fig:NLD_K1.5_colour}  obtained by the three LDs-based diagnostics, with respect to the one obtained by the SALI is $P_A \approx 94.4 \%$, $P_A \approx 92.5 \%$ and $P_A \approx 94.5 \%$ for $D_L^2$, $R_L^2$ and $S_L^2$ respectively. 

In Figs.~\ref{fig:NLD_K1.5_colour}(g)-(i) we respectively show for the $D_L^2$, $R_L^2$ and $S_L^2$ indices the regions in the considered $2$D subspace, where the indicators fail to correctly identify (with respect to the SALI categorization) the chaotic or regular nature of orbits. In particular, blue points correspond to regular (according to SALI) orbits which are falsely identified as chaotic by the $D_L^2$, $R_L^2$ and $S_L^2$ indicators, while red points denote orbits classified as chaotic by SALI, which are incorrectly identified as regular. Although the effectiveness of the three indicators in distinguishing between regular and chaotic orbits is clearly captured by the very high agreement percentages ($\gtrsim 90 \%$) with respect to the SALI classification, the results of  Figs.~\ref{fig:NLD_K1.5_colour}(g)-(i) show  that the large majority of incorrectly characterized orbits are mainly located at the edges of regular islands where sticky chaotic orbits exist, in agreement to what was reported in~\cite{Hillebrandchaos2022}. Our results show that the $D_L^2$ indicator falsely characterizes as regular many sticky chaotic orbits at the borders of stability islands [red points in Fig.~\ref{fig:NLD_K1.5_colour}(g)], while the use of $R_L^2$ and $S_L^2$ indices [Figs.~\ref{fig:NLD_K1.5_colour}(h), (i)] results in a more or less similar chart of wrongly identified orbits, which again are mainly located at the boarders of stability islands. It is worth noting that $S_L^2$  performs better than $R_L^2$  as it falsely characterizes as regular fewer chaotic orbits in the large chaotic sea [i.e.~there are fewer red points seen in the chaotic portion of Fig.~\ref{fig:NLD_K1.5_colour}(i) than in Fig.~\ref{fig:NLD_K1.5_colour}(h)].

\subsection{Effect of the number of iterations}
\label{sec:NLDs_int_time}

A key factor when studying chaotic systems is the integration time, or the number of iterations in the case of the $4$D map \eqref{eq:4Dmap},  required for indicators to correctly characterize orbits as regular or chaotic. In general, too few iterations do not allow for the exponential divergence of nearby orbits observed in the case of chaotic motion to lead to very large deviations, which in turn, would make apparent the chaotic nature of the orbits. This is true not only for indicators based on neighboring orbits' LDs but for any chaos indicator. On the other hand, too many iterations will make the use of the considered indicators less efficient as they will increase the required computational time. 

\begin{figure*}
        \begin{center}
        \includegraphics[width=0.99\textwidth]{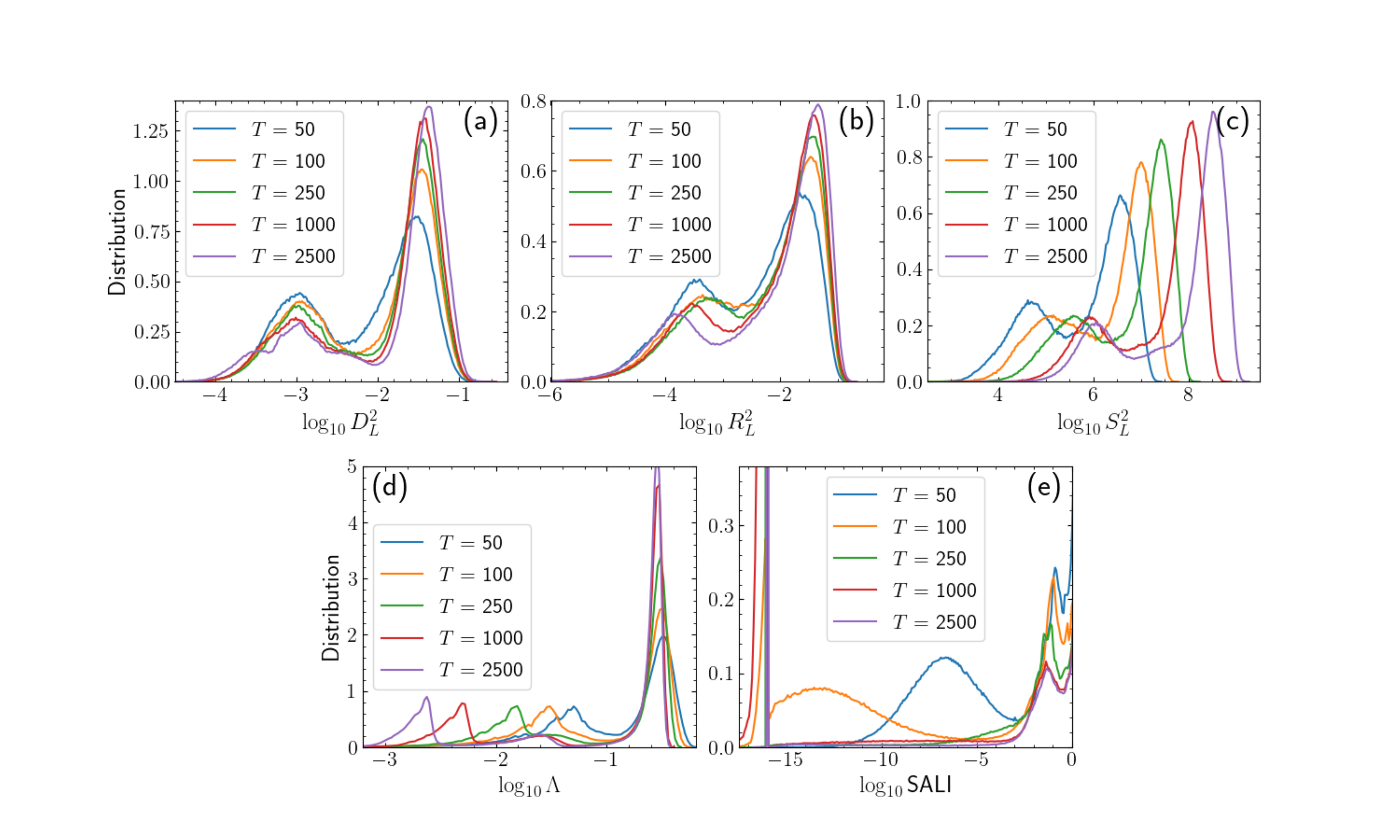}  
        \caption{Normalized distributions of the logarithms of the (a) $ D_L^2$ (\ref{eq:defDNLD}), (b) $ R_L^2$ (\ref{eq:defRNLD}), (c) $ S_L^2$ (\ref{eq:defDfNLD}), (d) $\Lambda$ (\ref{eq:mLE}), and (e) $ \mbox{SALI}$ (\ref{eq:SALI}), values of the orbits considered in Fig.~\ref{fig:NLD_K1.5_colour} for $T=50$ (blue curves), $T=100$ (orange curves), $T=250$ (green curves), $T=1000$ (red curves) and $T=2500$ (purple curves) forward iterations of the $4$D map \eqref{eq:4Dmap}.}
        \label{fig:Distri_compare}
        \end{center}
\end{figure*}

It is plausible to assume that the total number of iterations required for the characterization of orbits as chaotic or regular is directly related to the time it takes for the distributions of the $D_L^2$, $R_L^2$ and $S_L^2$ indices to clearly reveal two distinct peaks. When the two  peaks in the distribution are well formed, a threshold value can be  established between them allowing the discrimination between regular and chaotic orbits. Thus, in order to investigate the effect of the number of iterations $T$ on the behavior of the LDs-based diagnostics we respectively plot in Figs.~\ref{fig:Distri_compare}(a)-(c) the normalized distributions of the logarithms of the $D_L^2$, $R_L^2$ and $S_L^2$ values for the ensemble of orbits considered in Fig.~\ref{fig:NLD_K1.5_colour}. These distributions are computed for different numbers of forward iterations $T$ of map \eqref{eq:4Dmap}, namely for $T=50$ (blue curves), $T=100$ (orange curves), $T=250$ (green curves), $T=1000$ (red curves) and $T=2500$ (purple curves). From the results of these figures we see that the shape of the distribution of the three diagnostics does not significantly change, although in the case of $S_L^2$ [Fig.~\ref{fig:Distri_compare}(c)] the distribution is shifted towards larger $\log_{10} S_L^2$ values as $T$ increases, and that the distance between the peaks remains approximately constant. In addition,  for larger $T$ the height of the trough between the two well formed peaks decreases, allowing the more accurate characterization of the orbits' nature as it becomes easier to identify a well-placed threshold value between the two peaks. Nevertheless, since we would like to use the  $D_L^2$, $R_L^2$ and $S_L^2$ indices as a fast (i.e.~based on low iteration numbers) chaos indicator, we can say that, for the cases considered here, $T=1000$ is sufficient to properly capture the overall dynamics of the considered ensemble of orbits.   

For completeness' sake, we also present in Fig.~\ref{fig:Distri_compare} the evolution of the normalized distributions of the two basic chaos indicators we consider in our study, the ftmLE $\Lambda$  [Fig.~\ref{fig:Distri_compare}(d)] and the SALI [Fig.~\ref{fig:Distri_compare}(e)]. From Fig.~\ref{fig:Distri_compare}(d) we see that the distributions of the $\Lambda$ values have a high, sharp peak for $\log_{10} \Lambda \gtrsim -1$, which corresponds to the system's chaotic orbits for which $\Lambda$ eventually saturates to a positive value [see the orange curve in Fig.~\ref{fig:reg_vs_chaotic_plots}(b)]. In addition, we observe a second, smaller in this case, peak corresponding to regular orbits, which propagates to the left of Fig.~\ref{fig:Distri_compare}(d), towards smaller $\log_{10} \Lambda$ values, in agreement with Eq.~\eqref{eq:mLE_reg} [also see the blue curve in Fig.~\ref{fig:reg_vs_chaotic_plots}(b)]. The region between these two well formed peaks corresponds to weakly chaotic orbits for which  $\Lambda$ reaches  positive but small values.  On the other hand, the distribution of the SALI values [Fig.~\ref{fig:Distri_compare}(e)] develops  very fast, two well separated formations: a set of high positive values ($\log_{10} \mbox{SALI} \gtrsim -4$), which corresponds to regular orbits [see Eq.~\eqref{eq:SALI_4Dmap} and the blue curve in Fig.~\ref{fig:reg_vs_chaotic_plots}(c)], and a high peak at $\log_{10} \mbox{SALI} \approx -16$ corresponding to chaotic orbits whose SALI became practically zero reaching the level of the computer accuracy (i.e.~$10^{-16}$) due to the exponentially fast decrease of the index  [see Eq.~\eqref{eq:SALI_4Dmap} and the orange curve in Fig.~\ref{fig:reg_vs_chaotic_plots}(c)]. It is worth noting that even for as few iterations as $T=250$ [green curve in Fig.~\ref{fig:reg_vs_chaotic_plots}(c)] the distribution of the SALI values is practically flat and non-existing between the two well defined regions of small (chaotic orbits) and large (regular orbits) $\log_{10} \mbox{SALI}$ values. This fast distinction between the two categories of orbits is a main advantage of the SALI method, which also allows the establishment of a well defined threshold value for discriminating between regular and chaotic orbits, which in our work is set to $\log_{10} \mbox{SALI}=10^{-8}$. 

From the results of Fig.~\ref{fig:Distri_compare} we see that the increase of the number of iterations does not lead to a drastic improvement of the distinctive ability of the methods  based on neighboring orbits' LDs, as the shape of their distributions eventually does not change significantly [Figs.~\ref{fig:Distri_compare}(a)-(c)] in contrast to what happens with the distributions of the ftmLE and the SALI  shown in Figs.~\ref{fig:Distri_compare}(d) and \ref{fig:Distri_compare}(e). Thus, only slight adjustments are required for the threshold values of the $D_L^2$ and  $R_L^2$ indices for  the numbers of iterations reported in Fig.~\ref{fig:Distri_compare}. In contrast, a change in the number of iterations for $S_L^2$ results in the increase of the related threshold value. So, it is a good practice to check the value distributions of the three LDs-based quantities $D_L^2$, $R_L^2$ and $S_L^2$, in order to determine the optimal threshold values.

\subsection{Effect of the overall chaos extent and  grid spacing}
\label{sec:K_NLDs}

\begin{figure*}
    \begin{center}
    \includegraphics[width=0.99\textwidth]{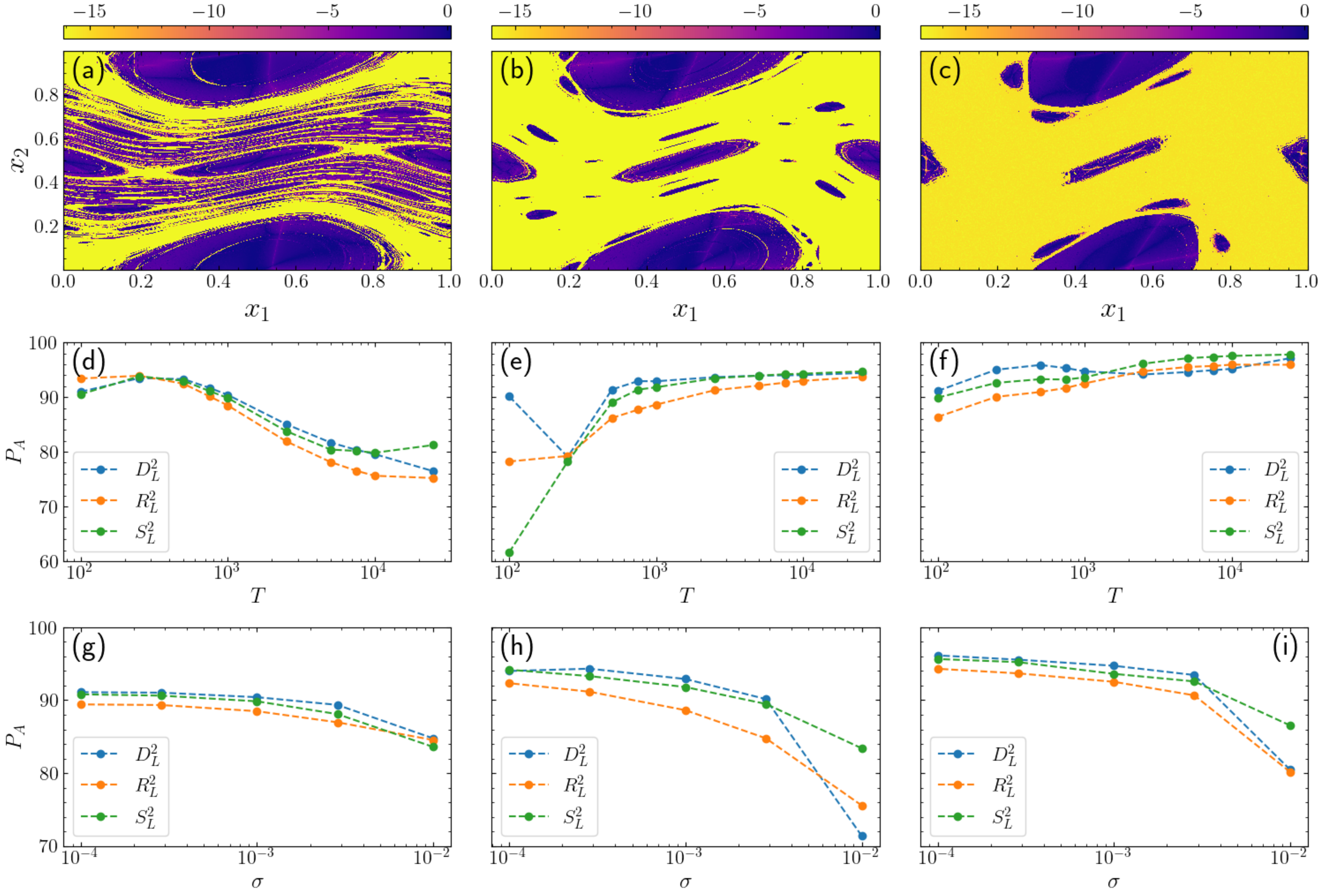}
    \caption{Results obtained for orbits having their ICs on a $1000 \times 1000$ grid on the $2$D subspace $(x_1, x_2)$ with $x_3 = 0.54$, $x_4 = 0.01$, of the $4$D map \eqref{eq:4Dmap} with $B=0.05$ and [(a), (d), (g)] $K=0.75$, [(b), (e), (h)] $K=1.1$,  [(c), (f), (i)] $K=1.5$. In (a)-(c) the ICs are colored according to the  orbits' $\log_{10} \mbox{SALI}$ value after $T=2.5 \times 10^4$ forward iterations using the color scales on the top of each panel. (d)-(f) The percentage accuracy $P_A$ of the orbits correctly characterized by the $D_L^2$ (blue points), $R_L^2$ (orange points) and $S_L^2$ (green points) with respect to the identification obtained by the SALI method for the same number of  iterations $T$, for the orbits respectively considered in (a)-(c). (g)-(i) The $P_A$ of orbits correctly characterized by the $D_L^2$, $R_L^2$ and $S_L^2$ (blue, orange and green points respectively) after $T=10^3$ iterations for five different grid spacings $\sigma$ on the $(x_1,x_2)$ space of (a)-(c) respectively. In all panels the $D_L^2$, $R_L^2$ and $S_L^2$ indices are evaluated through computations of neighboring orbits' LDs on the $2$D $(x_1,x_2)$ plane. In (d)-(i) the dashed line connections are used to guide the eye.}
    \label{fig:3SALI_Ks}
    \end{center} 
\end{figure*}

Let us now study  the effect of the system's chaoticity  on the accuracy of the indices. Both the nonlinearity parameter $K$ and the coupling constant $B$ of the $4$D map \eqref{eq:4Dmap} control the system's chaotic behavior, because, in general, their increase leads to more extended chaos. We investigated the performance of the three LDs-based diagnostics for various $K$ and $B$ values and we present here some representative results obtained by varying $K$, while $B$ is kept fixed. More specifically, in Figs.~\ref{fig:3SALI_Ks}(a)-(c), we show  SALI color plots for respectively  $K=0.75$, $K=1.1$ and $K=1.5$, and  $B=0.05$, computed for a total of $T = 2.5 \times 10^4$ forward iterations on a grid of $1000 \times 1000$ evenly spaced ICs on the $(x_1, x_2)$ plane with $x_3=0.54$ and $x_4=0.01$. From  these figures we clearly see that the increase of $K$ results in a substantial increase in the number of chaotic orbits, as the area of yellow-colored regions corresponding to very low SALI values (which indicate chaos) increases. In fact, we find the percentage $P_C$ of chaotic orbits to be $P_C \approx 43.9 \%$, $P_C \approx 69.8\%$ and $P_C \approx 79.6\%$ respectively for $K=0.75$, $K=1.1$ and $K=1.5$, when the $\log_{10} \mbox{SALI}=-8$ threshold is used to discriminate between regular and chaotic orbits. 

We next investigate, for the  three  cases of Fig.~\ref{fig:3SALI_Ks}, the effect of the total number of map iterations $T$ on the ability of the LDs-based indices to correctly capture the nature of orbits, which is quantified by their percentage agreement $P_A$ with the characterization obtained by SALI for the same $T$. We note that here we consider the order $n=2$ indices $D_L^2$, $R_L^2$ and $S_L^2$, which are based on LDs' computations of neighboring orbits on the  $2$D plane $(x_1, x_2)$ defined by $x_3 = 0.54$ and $x_4 = 0.01$ [Figs.~\ref{fig:3SALI_Ks}(a)-(c)]. The $P_A$ is computed for ten different final iteration numbers and the obtained results are presented in Figs.~\ref{fig:3SALI_Ks}(d)-(f). For each of the three considered $K$ values and the ten different final iteration numbers $T$  an appropriate threshold value for discriminating between regular and chaotic orbits is selected for every index following the approach described in Sect.~\ref{sec:NLDs_int_time}, while for the SALI the threshold value  $\log_{10}\mbox{SALI}=-8$ is always used. 

For $K=0.75$ [Figs.~\ref{fig:3SALI_Ks}(a) and \ref{fig:3SALI_Ks}(d)] the phase space displays the smallest area of chaotic behavior  among the three cases we considered, and $P_A$ decreases as the number of iterations increases. This is due to the large number of sticky orbits at the edges of the many regular islands, whose weakly chaotic nature is revealed by the SALI only after a rather high number of iterations. Thus, initially, for small $T$ values the SALI, as well as the $D_L^2$, $R_L^2$ and $S_L^2$ indices, wrongly characterize the sticky orbits as regular, but since all these methods agree on this assessment the related $P_A$ values in Fig.~\ref{fig:3SALI_Ks}(d) are large. For larger $T$ values the SALI eventually manages to identify the sticky obits as chaotic, but the LDs-based indicators fail to do so, and consequently the $P_A$ values decrease. This discrepancy is due to the known difficulty of the $D_L^2$, $R_L^2$ and $S_L^2$  indicators to correctly characterize sticky orbits, which has been already seen in Figs.~\ref{fig:NLD_K1.5_colour}(g)-(i). This limitation was also pointed out in~\cite{Hillebrandchaos2022}.  For $K=1.1$ [Figs.~\ref{fig:3SALI_Ks}(b) and \ref{fig:3SALI_Ks}(e)], the phase space's chaoticity increases and fewer sticky orbits are present compared to the $K=0.75$ case, and $P_A$ is observed to increase for large $T$ values, steadily exhibiting values $\gtrsim 90\%$. Similarly, for the highly chaotic case of $K=1.5$ [Figs.~\ref{fig:3SALI_Ks}(c) and \ref{fig:3SALI_Ks}(f)], for which the number of sticky orbits has been drastically reduced, as the extent of the chaotic sea has grown, $P_A$ enlarges with growing $T$. 

Our analysis shows again that  the $D_L^2$, $R_L^2$ and $S_L^2$ indicators are less efficient at properly characterizing orbits at the edges of regular regions.  This becomes especially problematic when the system's phase space is occupied by many stability islands and chaos is confined in  very thin strips between these islands, as is for example seen in the case  of Fig.~\ref{fig:3SALI_Ks}(a). Furthermore, the results of Figs.~\ref{fig:3SALI_Ks}(e) and \ref{fig:3SALI_Ks}(f) show that the $D_L^2$, $R_L^2$ and $S_L^2$ indices have similar chaos diagnostic capabilities as in almost all studied cases they achieve similar $P_A$ values for large enough $T$ numbers. In addition, taking also into account that we want to use these indicators as fast diagnostics, we observe that (as was also seen in Sect.~\ref{sec:NLDs_int_time})  $T=1000$ is a very good number for all indices to produces reliable estimations of chaos extent, especially for the $K=1.1$ and $K=1.5$ cases [Figs.~\ref{fig:3SALI_Ks}(e) and \ref{fig:3SALI_Ks}(f) respectively] for which $P_A \gtrsim 90\%$.

Having considered the effect of $T$ on the chaos diagnostic accuracy of $D_L^2$, $R_L^2$ and $S_L^2$, let us now discuss the effect of the grid spacing size $\sigma$ on their performance. For this purpose we respectively present in Figs.~\ref{fig:3SALI_Ks}(g)-(i) for $K=0.75$, $K=1.1$ and $K=1.5$ the $P_A$ values obtained by the three indicators at $T=10^3$ for five different grid spacings on the $(x_1, x_2)$ plane considered in Figs.~\ref{fig:3SALI_Ks}(a)-(c) in the range $10^{-4} \leq \sigma \leq 10^{-2}$. For each $K$ an increase in accuracy $P_A$ is seen as $\sigma$ decreases, indicating that computations based on finer grid capture  more accurately the system's dynamics. On the other hand, the use of more grid point results in a significant increase of the required computational time, something which is not desirable for the implementation of the $D_L^2$, $R_L^2$ and $S_L^2$ indices as fast chaos diagnostics. Nevertheless, the fact  that the accuracy $P_A$ for $\sigma = 10^{-4}$ is  only slightly better  than the one obtained for $\sigma = 10^{-3}$, suggests that after some point the further decrease of  the grid spacing  has only a moderate impact on the achieved accuracy. Thus, a rather good choice for the grid spacing in our study, balancing between the obtaining accuracy and the required computational effort, is $\sigma=10^{-3}$.

The results of Fig.~\ref{fig:3SALI_Ks} clearly show that the extent of chaos,  as well as its structure in the phase space, significantly influences the usefulness of the  $D_L^2$, $R_L^2$ and $S_L^2$ indicators  in characterizing the overall dynamics. In particular, we should be cautious when these indices are applied to  systems for which we  expect a small amount of chaos. Although this is a  limitation of the $D_L^2$, $R_L^2$ and $S_L^2$ indicators, it is worth noting that they still prove to be highly accurate in their characterization of orbits for systems with moderate or large $P_C$ values.

\subsection{Global dynamics and the role of the order of the LDs-based diagnostics}
\label{subsec:high_D_NLD}

So far we computed the $D_L^2$, $R_L^2$ and $S_L^2$ indices on $2$D subspaces  of the $4$D phase space of map \eqref{eq:4Dmap}. Now we will examine  what effect a change in the order $n$ of the three LDs-based indicators has  on their performance by considering not only their $n=2$ versions. As the order $n$ is increased we  are  adding and processing more information from the surroundings of a studied orbit, as we include in the evaluation of the $ D_L^n$, $ R_L^n$  and $ S_L^n$  indices the LD values of more neighboring orbits. Thus, we expect that the obtained results will capture more accurately  the nature of the underlying dynamics. Unfortunately, the increase of order $n$ comes with the drawback of the raised computational effort required to evaluate the LDs of the additional grid points used for evaluating the $D_L^n$, $R_L^n$ and $S_L^n$ indices. 

\begin{figure*}
        \begin{center}
        \includegraphics[width=0.99\textwidth]{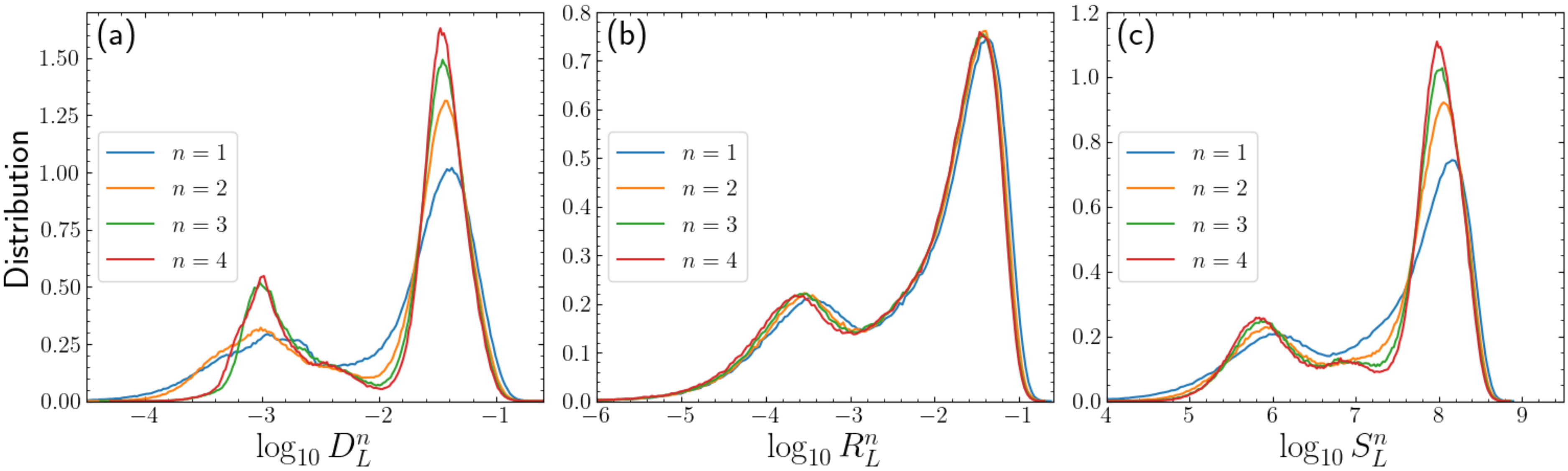}
        \caption{Normalized distributions of the (a) $\log_{10} D_L^n$ (\ref{eq:defDNLD}), (b) $ \log_{10} R_L^n$ (\ref{eq:defRNLD}), and (c) $\log_{10} S_L^n$, (\ref{eq:defDfNLD}) values of the orbits considered in Fig.~\ref{fig:NLD_K1.5_colour} for  orders $n=1$ (blue curves), $n=2$ (orange curves), $n=3$ (green curves) and $n=4$ (red curves). The $D_L^n$, $R_L^n$ and $S_L^n$ indices are evaluated along the $x_1$ direction for $n=1$, the $x_1$ and $x_2$ directions for $n=2$, the $x_1$, $x_2$ and $x_3$ directions for $n=3$ and all coordinate directions for $n=4$.}
        \label{fig:n_histo}
        \end{center}
\end{figure*}

The effect of order $n$ on the distributions of the $D_L^n$, $R_L^n$ and $S_L^n$ values is shown in Fig.~\ref{fig:n_histo} where these distributions are plotted for orders $n=1$ (blue curves), $n=2$ (orange curves), $n=3$ (green curves) and $n=4$ (red curves).  These distributions are obtained for the orbits considered in Fig.~\ref{fig:NLD_K1.5_colour}, whose ICs lie on the $2$D subspace $(x_1, x_2)$, $x_3=0.54$, $x_4=0.01$ of the $4$D map \eqref{eq:4Dmap} with $K=1.5$ and $B=0.05$. In particular, for $n=1$ neighboring orbits on the $x_1$ direction are considered for the computation of  $D_L^1$, $R_L^1$ and $S_L^1$, while for $n=2$ the nearby orbits are located on the $(x_1, x_2)$ plane. For the evaluation of the $D_L^3$, $R_L^3$ and $S_L^3$ indicators additional neighboring orbits with variations in their $x_3$ coordinates are considered, while orbits with variations also in the $x_4$ direction are used for the calculation of the order $n=4$ indices. We note that in all cases the grid spacing between neighboring orbits is $\sigma=10^{-3}$. 

From the results of Fig.~\ref{fig:n_histo} we see that for  the $D_L^n$ [Fig.~\ref{fig:n_histo}(a)] and $S_L^n$ distributions [Fig.~\ref{fig:n_histo}(c)] the two observed peaks increase in height as $n$ grows, although their positions do not change drastically, while at the same time the trough between them is decreasing. Thus, defining a threshold value for discriminating between regular and chaotic orbits becomes easier for larger $n$. It is also worth noting that the position of the threshold value at the minimum of the trough does not vary significantly with the indices' order, especially for $n \geq 2$. Interestingly, an increase in the order $n$ does not seem to have any effect on the shape of the distributions of the $R_L^n$ as shown in Fig.~\ref{fig:n_histo}(b). 

In order to gain a more general understanding on how the percentage accuracy $P_A$ of the three indicators changes with order $n$, and also to investigate the potential effect of the studied ensembles of orbits on the performance of the indices, the orbit classification obtained  by each indicator for $n=1, 2, 3$ and $4$ is compared to the SALI characterization for the same number of forward iterations, $T=10^3$, for six different sets of orbits. The examined ensembles of ICs are defined on the $2$D subspaces $(x_1,x_2)$, $(x_1,x_3)$, $(x_1, x_4)$, $(x_2,x_3)$, $(x_2,x_4)$ and $(x_3, x_4)$ of the $4$D map \eqref{eq:4Dmap} with $K=1.5$ and $B=0.05$, by considering a $1000\times1000$ evenly spaced grid of ICs (so that the grid spacing is $\sigma=10^{-3}$), while the remaining two variables are kept fixed at $x_1 = 0.6$, $x_2=0.2$, $x_3=0.54$, and $x_4=0.01$, depending on the $2$D subspace under consideration. The accuracy of each of the indicators $D_L^n$, $R_L^n$ and $S_L^n$, is then calculated for $ 1\leq n \leq 4$ for each set of ICs in the following way. For $n=1$ the indices are computed along the $x_i$ direction corresponding to the smaller $i$ index on the $2$D subspace, for $n=2$ along both directions of the $2$D subspace, while for $n=3$ the  $x_i$ direction  with the smaller $i$ index among the ones not included in the $2$D subspace is also considered. Obviously, for $n=4$ all directions are included in the computations. For example, in the case of the $(x_2,x_3)$ subspace the used ICs are on a $1000\times1000$  grid on the whole $(x_2,x_3)$ plane, i.e.~$0 \leq x_2 <1$, $0 \leq x_3 <1$, with $x_1 = 0.6$ and $x_4=0.01$. Then for $n=1$ the three indicators are computed by considering orbits along the $x_2$ direction,  for $n=2$ along both the $x_2$ and $x_3$ directions, and for $n=3$ along the $x_1$, $x_2$ and $x_3$ directions. The performed studies in the several subspaces, which cover a wide range of coordinate orientations, and for all the possible orders of the $D_L^n$, $R_L^n$ and $S_L^n$ indicators, ensure a global investigation of the indices' performance. The percentage $P_C$ of chaotic orbits for the six considered ensembles, according to the SALI classification for $T=10^3$, are   $P_C \approx 72.4\%$ for the $(x_1,x_2)$ case,  $P_C  \approx 91.8\%$ for $(x_1,x_3)$,   $P_C  \approx 89.6\%$ for $(x_1, x_4)$,  $P_C \approx 82\%$ for $(x_2,x_3)$,  $P_C \approx 77.7\%$ for $(x_2,x_4)$ and   $P_C \approx 84\%$ for the $(x_3, x_4)$ case.

\begin{figure*}
        \begin{center}
        \includegraphics[width=0.99\textwidth]{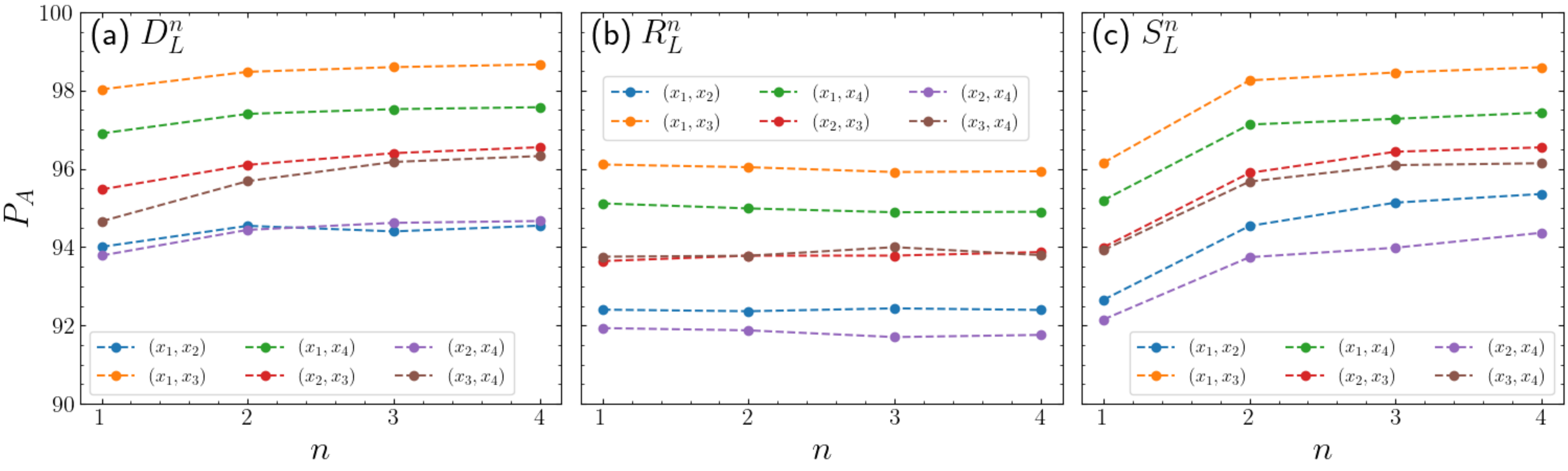}
        \caption{The percentage accuracy $P_A$ obtained by the (a) $ D_L^n$ (\ref{eq:defDNLD}), (b) $ R_L^n$ (\ref{eq:defRNLD}), and (c) $ S_L^n$ (\ref{eq:defDfNLD}) indices with respect to their order  $n$, for six different sets of ICs of the  $4$D  map \eqref{eq:4Dmap} with $K=1.5$ and $B=0.05$. The considered ensembles of orbits are defined on the $2$D subspaces $(x_1,x_2)$ (blue points), $(x_1,x_3)$ (orange points), $(x_1, x_4)$ (green points), $(x_2,x_3)$ (red points), $(x_2,x_4)$ (purple points) and $(x_3, x_4)$ (brown points)  by considering a $1000\times1000$  grid of ICs, while the remaining two variables are set to $x_1 = 0.6$, $x_2=0.2$, $x_3=0.54$, and $x_4=0.01$ depending on the $2$D subspace under consideration. The results are computed for $T=10^3$ forward iterations, and the dashed line connections are used to guide the eye. }
        \label{fig:P_A}
        \end{center}
\end{figure*}

In Fig.~\ref{fig:P_A} we present the percentage accuracy $P_A$ results obtained by the  $D_L^n$, $R_L^n$ and $S_L^n$ indices of order $1\leq n \leq 4$ for the six sets of considered ICs. From this figure we see that the efficiency of the $R_L^n$ index [Fig.~\ref{fig:P_A}(b)] in correctly capturing the regular or chaotic nature of the studied orbits does not practically depend on the order $n$, as for all considered cases its $P_A$ does not change with $n$. On the other hand, for the  $D_L^n$ [Fig.~\ref{fig:P_A}(a)] and the $S_L^n$ indices [Fig.~\ref{fig:P_A}(c)] we see a noticeable rise of  $P_A$ when $n$ is increased from $n=1$ to $n=2$ (which is more significant in the case of $S_L^n$), followed by a mild improvement as $n$ grows further. The main outcome of this analysis is that $n=2$ seems to be the optimal order for the three indicators, as setting $n>2$ does not result to significant improvements of the $P_A$ values, which would justify the associated increase  in the required computational time. We note that, due to the additional computations of LDs, the evaluation of indices of order $n=3$ ($n=4$) approximately requires three (six) times more computational effort with respect to the $n=2$ cases. 

\begin{figure}
        \centering
        \includegraphics[width=\columnwidth]{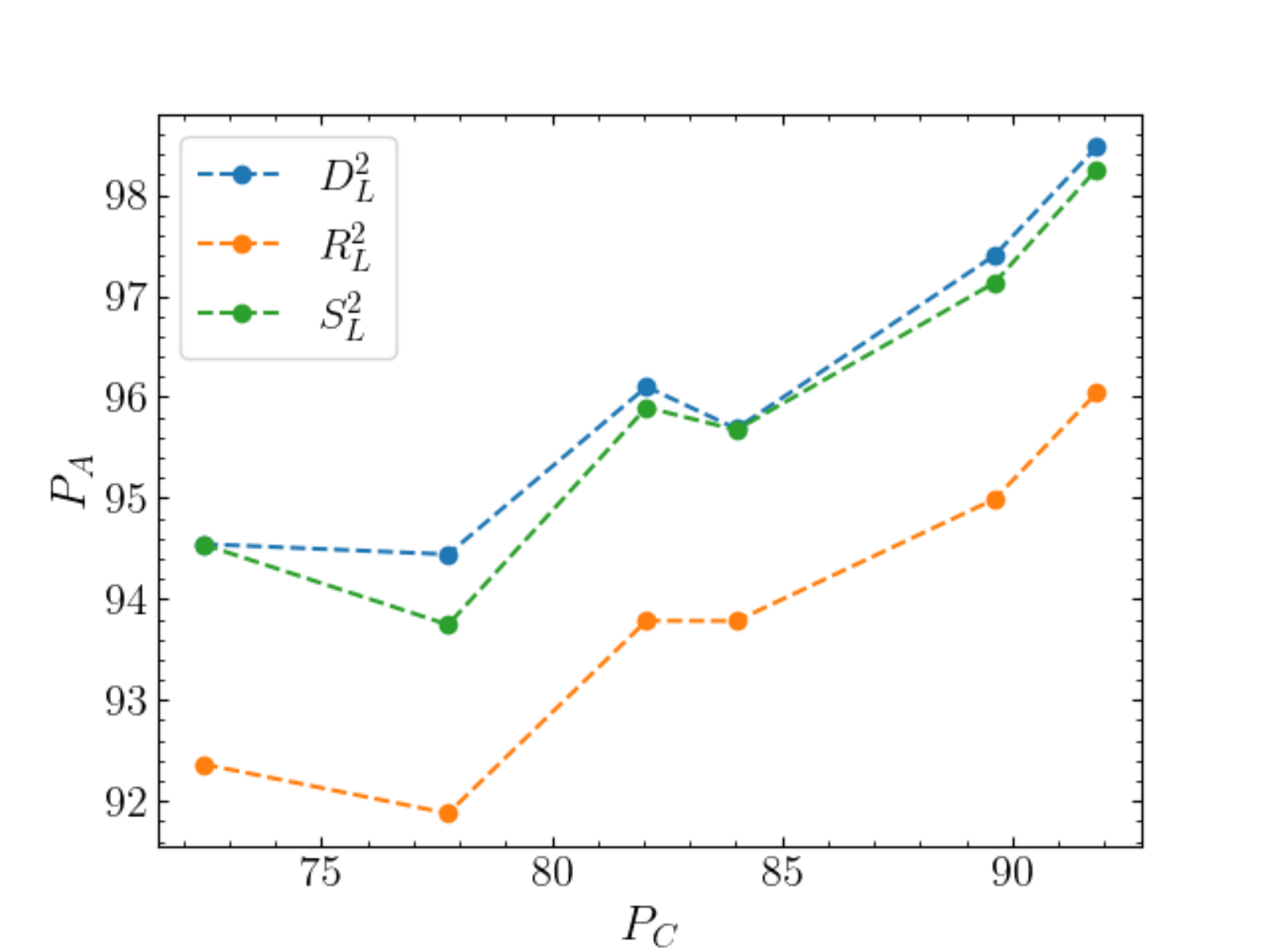}
        \caption{The percentage accuracy $P_A$ of the (a) $ D_L^2$ (\ref{eq:defDNLD}), (b) $ R_L^2$ (\ref{eq:defRNLD}), and (c) $ S_L^2$ (\ref{eq:defDfNLD}) indices  for the six different sets of ICs considered in Fig.~\ref{fig:P_A}, with respect to the  percentage $P_C$ of chaotic orbits  evaluated obtained by the SALI method. In all cases the related LDs and SALI values were calculated for a total of $T=10^3$ forward iterations. }
        \label{fig:P_C}
\end{figure}

In Fig.~\ref{fig:P_C} we see the percentage accuracy $P_A$ obtained by the $D_L^2$, $R_L^2$ and $S_L^2$ indices with respect to the percentage $P_C$ of chaotic orbits (obtained by the SALI method) in the six considered sets of ICs. As expected, a general increase in accuracy for the three indicators is seen as the percentage of chaos grows, with the $D_L^2$ and $S_L^2$ indices  being more accurate than $R_L^2$. This behavior  demonstrates again the fact that the three LDs-based indices become more accurate for more chaotic sets of orbits,  in accordance to the results discussed in Sect.~\ref{sec:K_NLDs}  [Fig.~\ref{fig:3SALI_Ks}].  

\begin{figure*}
        \begin{center}
        \includegraphics[width=0.99\textwidth]{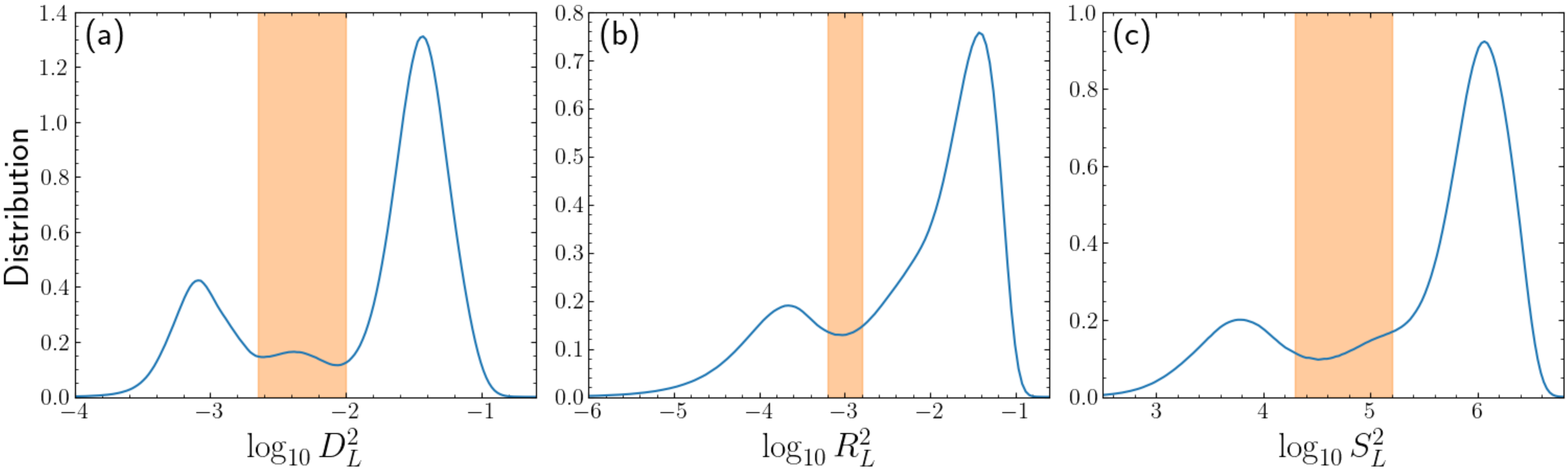}
        \caption{Normalized distributions of the (a)  $\log_{10} D_L^2$ (\ref{eq:defDNLD}), (b)  $\log_{10} R_L^2$  (\ref{eq:defRNLD}), and (c)  $\log_{10} S_L^2$ (\ref{eq:defDfNLD}) values of orbits with ICs in the intervals $x_1 \in [0.5,0.6)$, $x_2 \in [0,0.1)$,  $x_3 \in [0,1)$ and $x_4 \in [0,1)$, for the $4$D  map \eqref{eq:4Dmap} with $K=1.5$ and $B=0.05$, after $T=10^3$ forward iterations. In total $10^8$ ICs were considered on a grid having 100 points in each coordinate range. The $ D_L^2$, $ R_L^2$ and  $ S_L^2$ indices are evaluated on a $100 \times 100$ grid along the $x_1$ and $x_2$ directions  (which corresponds to a $\sigma=10^{-3}$ grid spacing). Taking different threshold values in the highlighted orange regions [(a) $-2.65 \leq \log_{10} D_L^2 \leq -2$, (b) $-3.2 \leq \log_{10} R_L^2 \leq  -2.8$ and (c) $4.3 \leq \log_{10} S_L^2 \leq 5.2$] alters the indices' percentage accuracy $P_A$ by about $2\%$. }
        \label{fig:4D_distri}
        \end{center}
\end{figure*}

As an additional example of the applicability of the three LDs-based diagnostics for investigating the global dynamics of map \eqref{eq:4Dmap}, we consider their implementation on a $4$D subspace of the system's phase space for $K=1.5$ and $B=0.05$. In particular we consider the subspace defined by $0.5 \leq x_1 <0.6$, $0 \leq x_2 <0.1$, $0 \leq x_3 <1$ and $0 \leq x_4 <1$, which corresponds to $1 \%$ of the total phase space. From the so far performed analyses we know that $n=2$ is the optimal order for achieving an accurate characterization of chaotic orbits and that LDs computations for $T=10^3$ forward iterations are sufficient for that purpose. Thus, we evaluate the $D_L^2$, $R_L^2$ and $S_L^2$ indices along two directions of the $4$D subspace, and in particular, along the $x_1$ and $x_2$ coordinates, by taking a grid of $100 \times 100$ points on the $(x_1,x_2)$ space, which corresponds to a $\sigma=10^{-3}$ grid spacing in accordance to the outcomes of Sect.~\ref{sec:K_NLDs}.  Furthermore, in order to get a good representation of the whole considered $4$D subspace, without unnecessarily increasing the number of studied ICs, we also regard a grid of $100 \times 100$ points along $x_3$ and $x_4$. This arrangement results in a total of $10^8$ ICs, with $P_C \approx 73 \%$ of them being chaotic according to their SALI values at $T = 10^3$. The resultant distributions of the $D_L^2$, $R_L^2$ and $S_L^2$ indices for this $4$D subspace are shown in Figs.~\ref{fig:4D_distri}(a)-(c) respectively, and have the same general shape as those seen in Figs.~\ref{fig:NLD_K1.5_colour}(d)-(f), Figs.~\ref{fig:Distri_compare}(a)-(c) and Fig.~\ref{fig:n_histo}, with two well-formed peaks corresponding to regular and chaotic orbits. The similarity of the obtained distributions for all considered cases in this work clearly indicates the generality of their shape, i.e.~two peaks with a trough in between, which defines the place of the indices' threshold value for identifying chaotic orbits. It is worth noting that the exact location of this threshold does not significantly alter the overall orbit characterization. In order to make this point more clear, for each distribution of Fig.~\ref{fig:4D_distri} we consider intervals for the location of the corresponding thresholds in the trough between the two peaks. These intervals are $-2.65 \leq \log_{10} D_L^2 \leq -2$, $-3.2 \leq \log_{10} R_L^2 \leq  -2.8$ and $4.3 \leq \log_{10} S_L^2 \leq 5.2$ and are denoted by the highlighted orange regions in each panel of  Fig.~\ref{fig:4D_distri}. Considering ten different evenly distributed threshold values in these intervals we found that the accuracy $P_A$ of the characterization made by the $D_L^2$, $R_L^2$ and $S_L^2$ indices, in comparison to one achieved  by the SALI, was in the ranges  $93.0 \% \lesssim P_A \lesssim 94.9 \%$ for the $D_L^2$ index, $91.8 \% \lesssim P_A \lesssim 92.6 \%$ for the $R_L^2$ indicator,  and $92.5 \% \lesssim P_A \lesssim 94.7 \%$ for the $S_L^2$ method. These results clearly illustrate that we can implement the $D_L^2$, $R_L^2$ and $S_L^2$ diagnostics  to distinguish between regular and chaotic orbits on a global scale in the phase space of the $4$D map \eqref{eq:4Dmap}, and that the selected threshold value does not have a strong impact on the accuracy of this characterization.

\section{Summary and Conclusion}
\label{sec:conclusions}

In this work, we investigated the ability of some simple quantities based on LDs computations to correctly identify orbits as regular or chaotic. In particular, we focused our attention on a conservative dynamical system whose phase space dimensionality makes the direct visualization of the dynamics a challenging task: the $4$D area preserving map \eqref{eq:4Dmap}, which is composed of two coupled $2$D standard maps. More specifically, the quantities  we considered were the difference $ D_L^n$ \eqref{eq:defDNLD}, and the ratio $ R_L^n$ \eqref{eq:defRNLD} of neighboring orbits' LDs, as well as the $ S_L^n$ index \eqref{eq:defDfNLD}, which is related to the second spatial derivative of the LDs. The     $ S_L^n$ index was initially presented in \cite{Daquin2022} (in a slightly different formulation to the one used in our study), while the $ D_L^n$ and $ R_L^n$ diagnostics were introduced in \cite{Hillebrandchaos2022}, where they were also applied to low-dimensional conservative dynamical systems, namely the two degree of freedom H\'enon--Heiles Hamiltonian and the $2$D standard map. Here, trying to investigate the applicability of these indices to high-dimensional systems, we considered a symplectic map having a $4$D phase space. We emphasize that all three indicators rely solely on computations of forward in time LDs (although backward LD computations produce similar results) of initially neighboring orbits, lying on $n$-dimensional spaces, with $n$ referred to as the order on each index. 

Although color plots of the $D_L^n$, $R_L^n$ and $S_L^n$ indices manage to correctly capture a qualitative picture of the system's dynamics [Figs.~\ref{fig:NLD_K1.5_colour}(a)-(c)], as also LDs themselves do, we showed that they can also, quite successfully, identify individual orbits as regular or chaotic, and consequently quantify the system's extent of chaos. Actually, in all studied setups the three LDs-based indices managed to correctly reveal the regular or chaotic nature of orbits with an agreement $P_A \gtrsim 90 \%$ with respect to the classification obtained by the SALI method. The importance of this achievement becomes  higher if we take into account the fact that the evaluation of the these indices depends only on the time evolution of orbits and does not require the knowledge of the related variational equations (in the case of continuous time systems) or the corresponding tangent map (for discrete time maps) governing the evolution of small perturbations to the studied orbits. 

In order to use the three LDs-based quantities as chaos diagnostics we defined appropriate threshold values from the distributions of the $D_L^n$, $R_L^n$ and $S_L^n$ indices [Figs.~\ref{fig:NLD_K1.5_colour}(d)-(f)]. These thresholds  were used to characterize an orbit as regular (chaotic) if its index value was below (above) the threshold. The determination of these thresholds was facilitated by the general shape of the  distributions, which have two well defined peaks, corresponding to chaotic (peak at higher index values) and  regular orbits (peak at lower values), separated by a trough [Figs.~\ref{fig:NLD_K1.5_colour}(d)-(f), Fig.~\ref{fig:n_histo}, and Fig.~\ref{fig:4D_distri}], where the threshold was set. Typically this threshold was defined at the distribution's minimum in the trough  [Figs.~\ref{fig:NLD_K1.5_colour}(d)-(f)], but the obtained orbit classifications were not too sensitive on the exact location of the threshold, as a variation of its value in the trough between the two peaks changed $P_A$ by $\lesssim 2 \%$  [Fig.~\ref{fig:4D_distri}].

Even though the general form of the $D_L^n$, $R_L^n$ and $S_L^n$  distributions  remained the same, their explicit shape and consequently the location of the threshold value for each index, depended on the number of iterations $T$ of the map for which the indices were computed [Figs.~\ref{fig:Distri_compare}(a)-(c)], as well as the order $n$ [Fig.~\ref{fig:n_histo}]. In general, the increase of $T$ and $n$ resulted in more pronounced peaks [with the exception of $R_L^n$ whose distribution does not seem to be affected by $n$; Fig.~\ref{fig:n_histo}(b)], while at the same time the trough's height  decreased making the determination of the threshold value easier, and the efficiency of the indices higher. Indeed an increase of $P_A$ was observed for various ensembles of studied orbits when $T$ [Figs.~\ref{fig:3SALI_Ks}(e) and (f)] and $n$ [Fig.~\ref{fig:P_A}] grew. On the other hand, the increase of $T$ and/or $n$ led to longer computations, as orbits were followed for more iterations when $T$ grew, and the number of neighboring orbits, whose LDs was needed for the indices' evaluation, increased for larger orders $n$. It is also worth noting that all distributions practically covered the  same value intervals, with the exception of the $S_L^n$ which was shifted to higher index values when $T$ increased [Fig.~\ref{fig:Distri_compare}(c)]. Trying to find a balance between the achieved accuracy $P_A$ in identifying chaos and the overall required computational time, in order to use the $D_L^n$, $R_L^n$ and $S_L^n$ indices as efficient, short time chaos diagnostics, we showed that good choice for the $T$ and $n$ variables are $T=1000$ and $n=2$. Another factor which influenced the accuracy and the efficiency of the three indicators was the initial phase space distance (grid spacing $\sigma$) between the neighboring orbits for which LDs were computed. We showed that a finer grid (smaller distances) led to more accurate results and higher $P_A$ values [Figs.~\ref{fig:3SALI_Ks}(g)-(i)], having at the same time the obvious drawback of the increase of required computational effort as more orbits were evolved. Our analysis indicated that a good balance between these two factors was obtained for $\sigma=10^{-3}$. 

We also explored the effect on the performance of the three indicators of the system's extent of chaos, i.e.~the fraction $P_C$ of chaotic orbits, as this was defined by the SALI method. Our results showed  that the indicators perform better for systems with higher $P_C$ values. More specifically, we found that the three diagnostics mainly failed to correctly identify the nature of orbits located at the edges of stability islands, where sticky chaotic orbits exist [Figs.~\ref{fig:NLD_K1.5_colour}(g)-(i)]. Consequently, the efficiency of these indices was decreased when the system's phase space was occupied by many stability islands of various sizes, having narrow chaotic strips between them where many  sticky orbits resided [Figs.~\ref{fig:3SALI_Ks}(a), (d) and (g)]. Nevertheless, even in such cases, an appropriate selection of the computation variables (in our case $n=2$, $T=1000$ and $\sigma=10^{-3}$) led to good results with $P_A \gtrsim 90 \%$. The main outcome of that investigation is that a fair amount of care should be taken for application of these LDs-based indices to systems where low levels of chaos are expected. 

In summary, we found that, with respect to the variations of the distributions of the different indices (which affect the determination of the threshold value for discriminating  between  regular and chaotic orbits), the $S_L^n$ distributions were significantly affected (moved to higher values) as $T$ grew, although they more or less retained their shape [Fig.~\ref{fig:Distri_compare}(c)]. On the other extreme end, the $R_L^n$ distributions were not influenced by order $n$ [Fig.~\ref{fig:n_histo}(b)]. In all other cases we observed slight distribution variations with respect to $T$ [Figs.~\ref{fig:Distri_compare}(a) and (b)] and $n$  [Fig.~\ref{fig:n_histo}(a) and (c)], which  led to small (if any) changes in the considered threshold values, that nevertheless did not drastically affect the overall orbit classification [Fig.~\ref{fig:4D_distri}]. 

From the results presented in this study, it is apparent that, in general, the $D_L^n$ and $S_L^n$ indicators performed  better than  $R_L^n$ as they achieved larger $P_A$ values [Figs.~\ref{fig:3SALI_Ks}(d)-(i) and  \ref{fig:P_C}]. Thus, if only one index is to be used for the  global investigation of the chaotic behavior of a model, we recommend this indicator to be $D_L^n$ or $S_L^n$, with, in general, the latter being a preferable choice as it performed slightly better with respect to the obtained $P_A$ values  [Figs.~\ref{fig:3SALI_Ks}(d)-(i) and  \ref{fig:P_C}], although its threshold value significantly varies with $T$ [Fig.~\ref{fig:Distri_compare}]. Nevertheless, once the LDs have been computed for a tested ensemble of orbits, evaluating any of the three indicators  is a straightforward task. It is worth noting that although the results obtained by  the $D_L^n$ and $S_L^n$ indicators are not as precise as those achieved by  standard chaos detection techniques like the   SALI, the computations needed for their evaluation do not require the knowledge of the variational equations or the construction of the related tangent map, which simplifies the process of revealing the chaoticity of orbits.

We emphasize that the generality of our outcomes is supported by the fact that the presented results were obtained for several sets of ICs located in various subspaces of the map's phase space, having different dimensions, and for different parameter values. Our findings show that tools based on LDs computations can be effectively used as chaos diagnostic techniques also for conservative dynamical systems of higher dimensions, extending and completing in this way the results presented in \cite{Hillebrandchaos2022}.

\section*{Declaration of competing interest}
The authors declare that they have no known competing financial interests or personal relationships that could have appeared to influence the work reported in this paper.

\section*{Acknowledgments}
A.~N.~acknowledges support from the University of Cape Town (University Research Council, URC) postdoctoral Fellowship grant and  from the Oppenheimer Memorial Trust (OMT). M.~H.~acknowledges support by the National Research Foundation (NRF) of South Africa (grant number 129630). M.~K.~and S~.W.~acknowledge the financial support provided by the EPSRC Grant No. EP/P021123/1. We thank the High Performance Computing facility of the University of Cape Town and the Centre for High Performance Computing \cite{chpc} of South Africa for providing computational resources for this project.

\bibliographystyle{elsarticle-num} 
\bibliography{ZNHKWS}

\end{document}